\documentclass[aps,prl,superscriptaddress,amsfonts,amsmath,amssymb,reprint,showpacs,floatfix]{revtex4-1}

\usepackage{url}
\usepackage{bm}
\usepackage{graphicx}
\usepackage{amsmath}
\usepackage{amstext}
\usepackage{amssymb}
\usepackage{amsfonts}
\usepackage{amsbsy}
\usepackage{verbatim}
\usepackage{color}
\usepackage[colorlinks=true, urlcolor=blue, linkcolor=blue, citecolor=blue, pdftex]{hyperref}
\usepackage{multirow}
\usepackage{pdfpages}
\usepackage{floatrow}
\usepackage{gensymb}
\usepackage{textcomp}
\usepackage{ulem}

\newcommand{\bs}[1]{\boldsymbol{#1}}

\begin{document}

\title{Functional renormalization group for three-dimensional quantum magnetism}

\author{Yasir Iqbal}
\email[]{yiqbal@physik.uni-wuerzburg.de}
\affiliation{Institute for Theoretical Physics and Astrophysics, Julius-Maximilians University of W\"urzburg, Am Hubland, D-97074 W\"urzburg, Germany}
\author{Ronny Thomale}
\affiliation{Institute for Theoretical Physics and Astrophysics, Julius-Maximilians University of W\"urzburg, Am Hubland, D-97074 W\"urzburg, Germany}
\author{Francesco Parisen Toldin}
\affiliation{Institute for Theoretical Physics and Astrophysics, Julius-Maximilians University of W\"urzburg, Am Hubland, D-97074 W\"urzburg, Germany}
\author{Stephan Rachel}
\affiliation{Institut f\"ur Theoretische Physik, Technische Universit\"at Dresden, D-01062 Dresden, Germany}
\author{Johannes Reuther}
\affiliation{Dahlem Center for Complex Quantum Systems and Fachbereich Physik, Freie Universit\"at Berlin, D-14195 Berlin, Germany}
\affiliation{Helmholtz-Zentrum Berlin f\"ur Materialien und Energie, D-14109 Berlin, Germany}

\date{\today}

\begin{abstract}
We formulate a pseudofermion functional renormalization group (PFFRG) scheme to address frustrated quantum magnetism in three dimensions. In a scenario where many numerical approaches fail due to sign problem or small system size, three-dimensional (3D) PFFRG allows for a quantitative investigation of the quantum spin problem and its observables. We illustrate 3D PFFRG for the simple cubic $J_1$-$J_2$-$J_3$ quantum Heisenberg antiferromagnet, and benchmark it against other approaches, if available. 
\end{abstract}

\maketitle

{\it Introduction.} Frustrated quantum magnetism has established broad experimental and theoretical interest in condensed matter~\cite{frust,diep}. In particular, from the viewpoint of quantum paramagnets as potential
hosts of unconventional quantum states of matter~\cite{Anderson-1973,Balents-2010}, this field has persisted until today, and keeps generating manifold connections to other areas of contemporary research such as topological phases and quantum information. 

From a methodological perspective, the microscopic investigation of three-dimensional (3D) frustrated quantum magnetism constitutes one of the biggest challenges, which to a large extent remains unresolved. Mean-field approaches for quantum magnetism such as Schwinger bosons~\cite{PhysRevB.38.316}, along with spin waves, and linked cluster expansions~\cite{seriesexp} are often efficient to describe magnetic order in 3D but tend not to accurately capture paramagnetic behavior. While density-matrix renormalization group (DMRG)~\cite{RevModPhys.77.259} is the method of choice for one-dimensional spin systems, and extensions to two dimensions (2D) have proven useful in many cases~\cite{whity}, applications in 3D are unfeasible due to system size and entanglement scaling. The application of variational Monte Carlo (VMC)~\cite{PhysRev.138.A442,PhysRevB.16.3081} methods, equipped with an efficient mean-field description of magnetic and paramagnetic states including spin liquids~\cite{PhysRevB.65.165113}, has likewise been predominantly constrained to 2D~\cite{PhysRevB.87.060405}: While an increase in the number of wave function parameters to be optimized is in principle no issue, VMC still suffers from system size limitation when computing expectation values of observables. Whereas sufficient system size can be reached by quantum Monte Carlo approaches~\cite{PhysRevB.37.5978,PhysRevB.43.5950}, they are constrained to bipartite lattices with nonfrustrating spin interactions, and as such mostly do not allow access to the domains of interest.  

\begin{figure*}[t]
\includegraphics[width=1.0\columnwidth]{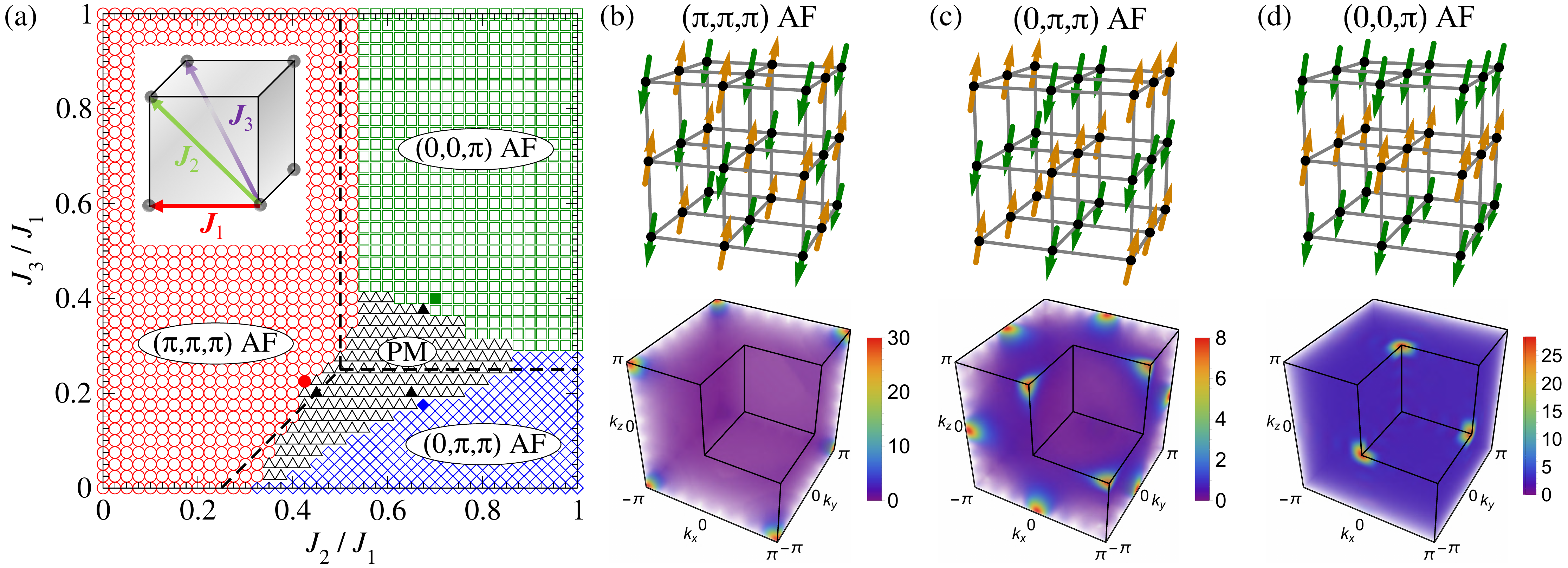}
\caption{\label{fig:fig1}
(Color online) (a) Quantum phase diagram of the spin-$\frac{1}{2}$ simple cubic $J_1$-$J_2$-$J_3$ Heisenberg antiferromagnet [see Eq.~(\ref{ham})] as obtained by PFFRG. It features a paramagnetic domain surrounded by commensurate magnetic orders. (Classical magnetic phase boundaries are drawn by black dashed lines for comparison.) (b)-(d) Illustrations of the real space pattern (upper row) and magnetic susceptibility profile (lower row) in units of $1/J_1$ for magnetism at ordering vectors $\bs{Q}=(\pi,\pi,\pi),(0,\pi,\pi),$ and $(0,0,\pi)$ obtained for the parameters $(J_2/J_1,J_3/J_1)=(0,0),(1,0),$ and $(1,1)$, respectively.}
\end{figure*}

In this Rapid Communication, we propose a pseudofermion functional renormalization group (PFFRG) scheme to describe frustrated quantum magnetism in 3D. While methodologically the 3D PFFRG is similar to previous formulations in 2D~\cite{Reuther-2010,Reuther-2011a}, it remedies some shortcomings of 2D PFFRG, allowing for a more accurate analysis of quantum magnetism. To illustrate the 3D PFFRG, we investigate the spin-$\frac{1}{2}$ $J_1$-$J_2$-$J_3$ Heisenberg antiferromagnet on the simple cubic lattice (SC-AFM), the ground-state phase diagram of which is summarized in Fig.~\ref{fig:fig1}(a). The parallelizability of the renormalization group (RG) flow equations guarantees accessibility to system sizes greater than $4000$ sites, which is an order of magnitude beyond other numerical methods available. Furthermore, in contrast to 2D, where the Mermin-Wagner theorem only allows for magnetic order at $T=0$, the finite ordering scales which naturally occur in PFFRG due to hidden mean-field character, can now be directly interpreted as ordering temperatures. In fact, wherever we are able to compare due to the absence of a sign problem, we find quantum Monte Carlo results in remarkable quantitative agreement with 3D PFFRG. Furthermore, the momentum-resolved spin correlations from 3D PFFRG, due to the large system sizes available, allow one to make contact with experimental observables, rendering it a promising approach in frustrated quantum magnetism. 

{\it Pseudofermion FRG.} Given a lattice on which spin-$1/2$ degrees of freedom are defined that interact through some spin Hamiltonian, the starting point of the PFFRG is to express the spin operators in terms of auxiliary fermions~\cite{Abrikosov-1965} $S_i^\mu=\frac{1}{2}\sum_{\alpha \beta} f_{i\alpha}^\dagger \sigma_{\alpha \beta}^\mu f_{i\beta}^{\phantom{\dagger}}$. Here, $f_{i\uparrow}$ ($f_{i\downarrow}$) denotes a fermionic annihilation operator of spin $\uparrow$ ($\downarrow$) on site $i$, and $\sigma^{\mu}$, $\mu \in \{x, y, z\}$ denote the Pauli matrices. According to standard procedures in FRG \cite{Metzner-2012,Platt-2013}, a frequency cutoff $\Lambda$ is implanted into the generating functional of vertex functions to yield an infinite set of coupled flow equations for all many-particle vertex functions. Along with accounting for the artificial enlargement of the pseudofermion Hilbert space by projection onto single occupancy per site, any bilinear spin Hamiltonian maps to a quartic pseudofermion interaction which constitutes the initial condition of the RG flow at $\Lambda \to \infty$. In a numerical implementation, the hierarchy of RG equations is truncated and only the self-energy and the two-particle vertex functions are kept. Most importantly, this truncation is performed such that the PFFRG remains separately exact in the limits of large $S$ (the magnitude of the spins) and large $N$ [where $N$ generalizes the spin symmetry group from $SU(2)$ to $SU(N)$]. Three-particle terms that are subleading in $1/S$ and $1/N$ are neglected. While the leading $1/S$ terms (i.e., spin mean-field terms) reproduce the classical magnetic phases, the $1/N$ terms add quantum fluctuations to the system and allow for the formation of paramagnetic phases. The physical outcome of the PFFRG is the static spin-spin correlator as a function of $\Lambda$ which is derived from the two-particle vertex. The flow equations in 3D remain invariant compared to their previous form in 2D PFFRG \cite{Reuther-2010,Reuther-2011a}. If a system adopts magnetic order, the corresponding two-particle vertex channel anomalously grows under the RG flow and eventually causes the flow to become unstable as the channel flows towards strong coupling.

\begin{figure}[b]
\includegraphics[width=1.0\columnwidth]{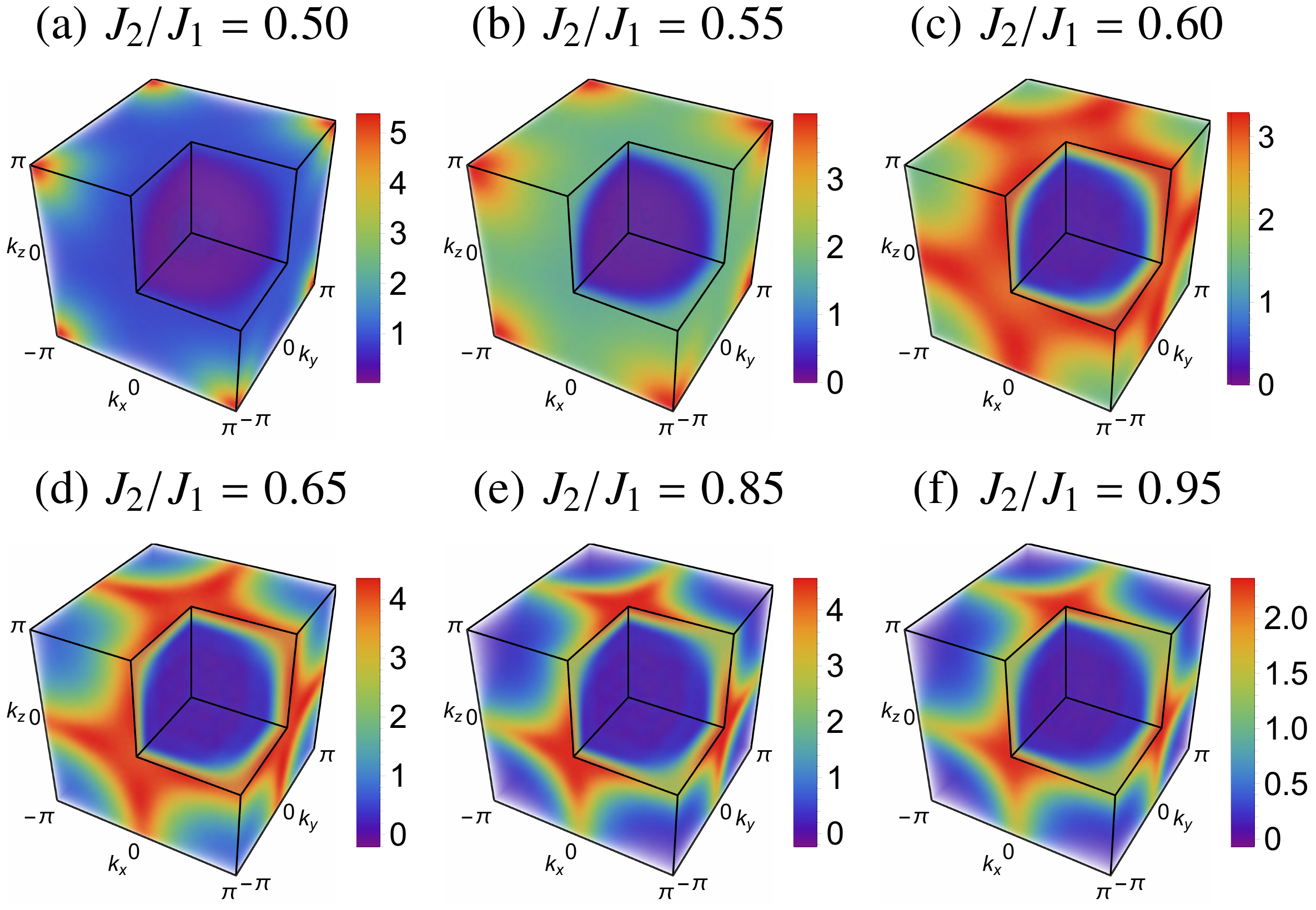}
\caption{\label{fig:fig2}
(Color online) (a)-(f) Evolution of magnetic susceptibility for $J_3/J_1=0.3$ from $J_2/J_1=0.50$ to $0.95$. As we traverse the paramagnetic region into the $(0,0,\pi)$ antiferromagnet (AF) in close proximity to the $(0,\pi,\pi)$ AF for lower $J_3/J_1$, the short-range correlations deviate significantly from the characteristic susceptibility profile deep in the ordered regimes [see Figs.~\ref{fig:fig1}(b)-\ref{fig:fig1}(d)]. We have cut out one octant of the Brillouin zone such that one can also visually track the spin correlations around the $\Gamma$ point.}
\end{figure}

\begin{figure*}[t]
\includegraphics[width=1.0\columnwidth]{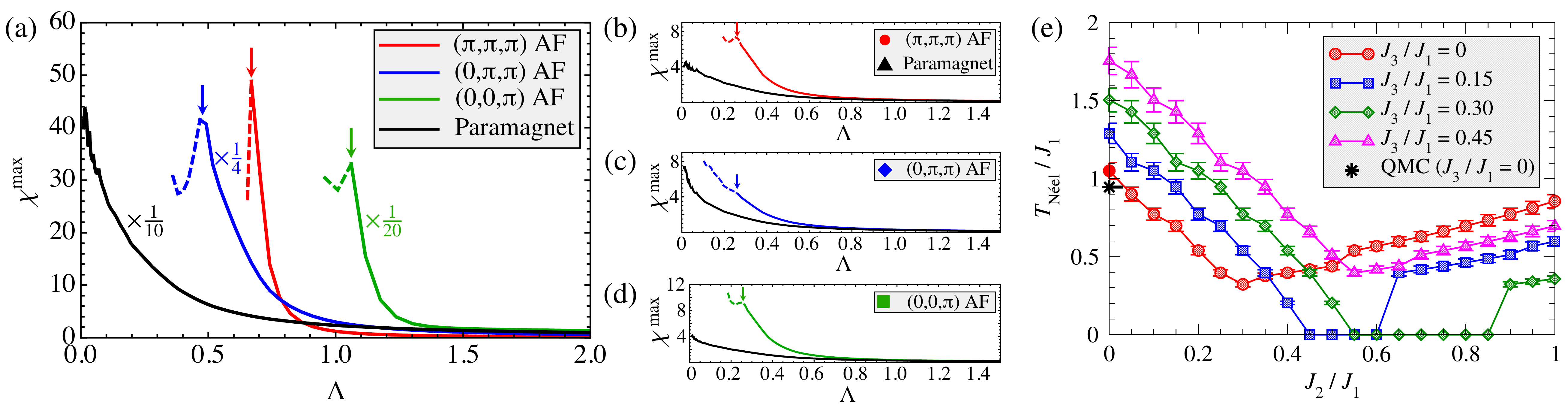}
\caption{\label{fig:fig3}
(Color online) (a) PFFRG coupling flow deep in the paramagnetic and different magnetic phases, shown for the parameter values $(J_{2}/J_{1},J_{3}/J_{1})=(0.5,0.25)~[{\rm Paramagnet}]$, $(0,0)~[(\pi,\pi,\pi)]$, $(1,0)~[(0,\pi,\pi)]$, and $(1,1)~[(0,0,\pi)]$. In the case of ordering, the flow exhibits a sharp singularity at $\Lambda^* \equiv \frac{2}{\pi}T_{\text{N\'eel}}$. (b)-(d) Distinct change in flow behavior from the paramagnet into the magnetic phases highlighted by pairs
of filled data point labels in Fig.~\ref{fig:fig1}(a). (e) $T_{\text{N\'eel}}$ for different cuts of constant $J_3/J_1$ as a function of $J_2/J_1$. The black asterisk highlights the value for the $J_1$ model as obtained by quantum Monte Carlo~\cite{PhysRevLett.80.5196}.}
\end{figure*}

Since the RG parameter $\Lambda$ and the temperature $T$ both act like a low energy cutoff, one can interpret the flow $\Lambda \to 0$ as an effective annealing process where the scale $\Lambda$ is associated with $\frac{2}{\pi}T$ \footnote{The conversion factor between the temperature $T$ and the RG scale $\Lambda$ can be found by comparing the limit of PFFRG where only the RPA diagrams contribute which corresponds to a mean-field description, and the conventional spin mean-field theory formulated in terms of temperature instead of $\Lambda$}. 2D PFFRG has proven capable of describing various phenomena of frustrated magnetism that are in principle complicated to treat, such as incommensurate spiral order with large ordering vectors~\cite{Reuther-2011b,PhysRevB.90.100405}, quantum order by disorder~\cite{PhysRevX.5.041035}, finite temperature behavior~\cite{Reuther-2011c}, anisotropic spin interactions~\cite{PhysRevLett.108.127203, PhysRevB.85.214406}, and strong geometric frustration~\cite{Suttner-2014,Iqbal-2015c,*Iqbal-2016a}. On the other hand, due to numerical frequency discretization and flow equation truncation, the 2D PFFRG erroneously gives finite ordering scales $\Lambda^*$ even though continuous symmetries do not allow for their spontaneous breaking at finite temperature~\cite{PhysRevLett.17.1133}. Turning to 3D where the Mermin-Wagner theorem allows for magnetic order at finite temperatures, the PFFRG ordering scales are directly linked to the N\'eel temperature via $T_\text{N\'eel}=\frac{\pi}{2}\Lambda^*$. At the same time, the appreciable features from 2D directly carry over to 3D PFFRG, such as the absence of a sign problem, resolvability of spin correlations as a function of (effective) temperature, exact correlations at $S\to\infty$ and $N\to\infty$, or the accessibility of large system sizes. In particular, the latter guarantees good momentum resolution of spin susceptibilities [see Figs.~\ref{fig:fig1}(b)-\ref{fig:fig1}(d) and Fig.~\ref{fig:fig2}], which is vital to making a comparison with, e.g., neutron scattering data~\cite{Suttner-2014}. 

{\it $J_1$-$J_2$-$J_3$ simple cubic antiferromagnet.}
 The SC-AFM  

\begin{equation}
\mathcal{H}_{\text{sc}}= J_1\sum_{\langle i,j \rangle} \mathbf{S}_i \cdot \mathbf{S}_j + 
J_2 \sum_{\langle \langle i,j \rangle \rangle } \mathbf{S}_i \cdot \mathbf{S}_j + J_3 \sum_{\langle
  \langle \langle i,j  \rangle \rangle \rangle } \mathbf{S}_i \cdot \mathbf{S}_j 
\label{ham}
\end{equation}

comprises antiferromagnetic Heisenberg coupling between first-~($J_1$), second-~($J_2$), and third-nearest neighbors ($J_3$). Exploiting lattice symmetries and massive parallelization, we solve the PFFRG flow equations for a linear cubic size up to $L=17$, totalling 4913 sites. The ground-state phase diagram [see Fig.~\ref{fig:fig1}(a)] features three magnetic regimes of $(\pi,\pi,\pi)$, $(\pi,\pi,0)$, and $(\pi,0,0)$ orders [see Figs.~\ref{fig:fig1}(b)-\ref{fig:fig1}(d)] as well as a paramagnetic (PM) domain setting in for finite $J_3$. As compared to the classical boundaries marked by dashed black lines in Fig.~\ref{fig:fig1}(a), quantum fluctuations hardly affect the $(\pi,\pi,\pi)$ regime while the paramagnet predominantly settles in within the classical domains of the stripe/plane collinear orders. Deep inside the magnetically ordered phases, the susceptibility is strongly peaked at the respective classical ordering wave vectors [see Figs.~\ref{fig:fig1}(b)-\ref{fig:fig1}(d)]. Figure~\ref{fig:fig2} depicts momentum-resolved spin susceptibilities (this quantity is computed at $\Lambda=0$ when possible or just before the possible breakdown when some momentum shows an instability) for a cut along $J_2$ for fixed $J_3/J_1=0.3$. Figure~\ref{fig:fig2}(a) starts off where $(\pi,\pi,\pi)$ order is still present. As we enter the PM phase, the susceptibility suddenly spreads out in momentum space and takes on profiles that qualitatively differ from those in the magnetically ordered domains. For large $J_2$, the susceptibility profile eventually merges into $(0,0,\pi)$ order, however, subleading features at $(0,\pi,\pi)$ originating from the nearby $\mathbf{Q}=(0,\pi,\pi)$ phase are still noticeable.

\begin{figure}[b]
\includegraphics[width=1.0\columnwidth]{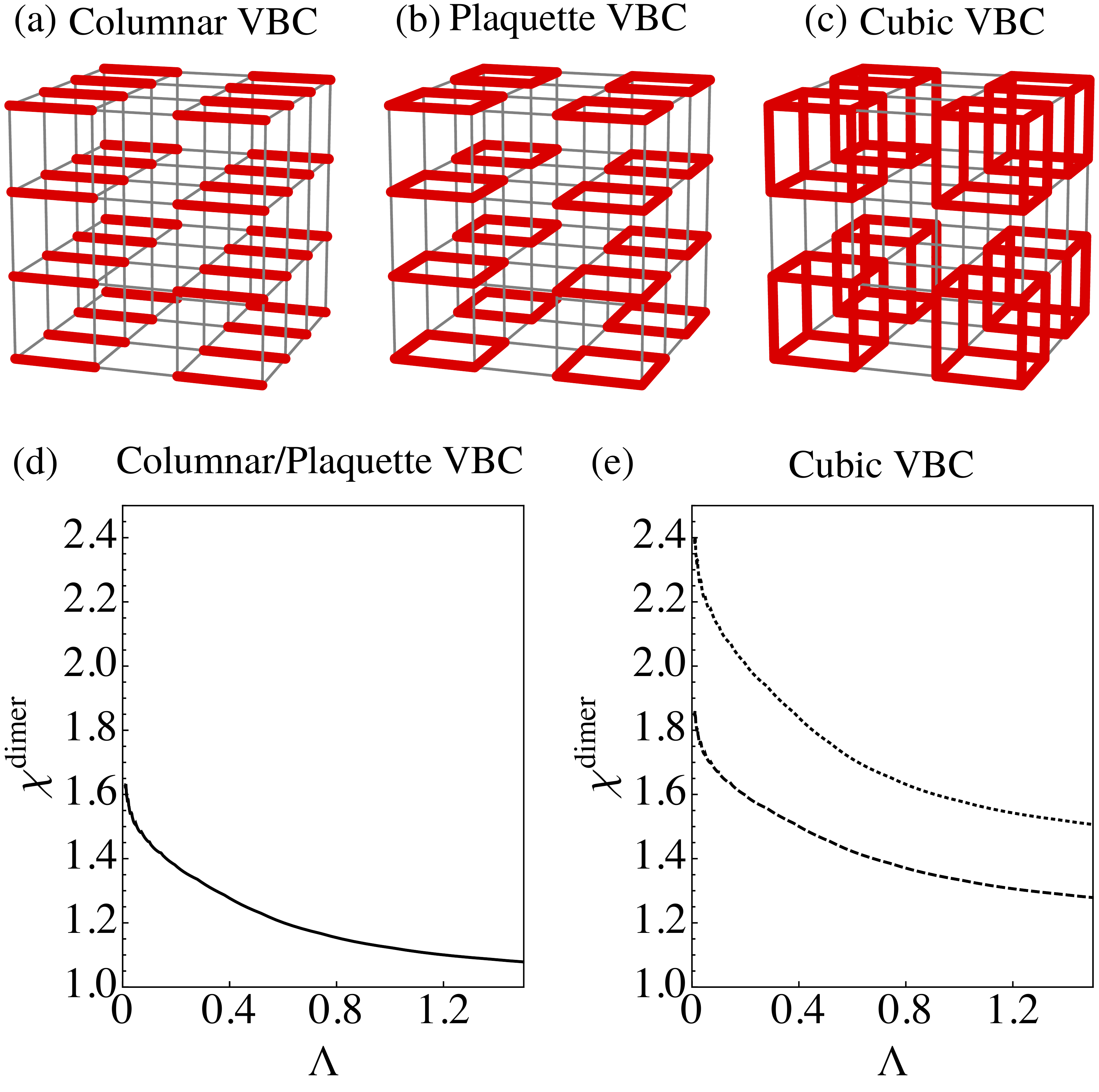}
\caption{\label{fig:fig4}
(Color online) (a)-(c) Valence bond crystal candidates in the paramagnetic regime of Fig.~\ref{fig:fig1}. Red bonds correspond to strengthened ($J_1\rightarrow J_1+\delta$) and gray bonds to weakened ($J_1\rightarrow J_1-\delta$) nearest-neighbor interactions. (d) Columnar and plaquette VBC show nearly identical and weak dimer response strength throughout the paramagnetic regime (variations of the response within the PM phase are smaller than the thickness of the line). (e) The cubic VBC exhibits a similarly weak
response strength which, however, varies as a function of $J_2/J_1$ and $J_3/J_1$, shown here for $(J_{2}/J_{1},J_{3}/J_{1})=(0.45,0.15)$ (dotted line) and $(0.70,0.25)$ (dashed line).}
\end{figure}

Figure~\ref{fig:fig3}(a) shows characteristic vertex flow evolutions towards magnetic order and paramagnetism. As compared to 2D PFFRG, we notice a much sharper instability behavior which strongly reduces the error imposed on distinguishing weak order from paramagnetism and on specifying $T_{\text{N\'eel}}$. To illustrate the former, we have taken pairs of points on opposite sides of a phase transition line between magnetic order and the PM phase [indicated by filled symbols in Fig.~\ref{fig:fig1}(a)], and show how strongly the flow evolution changes between them [see Figs.~\ref{fig:fig3}(b)-\ref{fig:fig3}(d)]. While there is a particularly drastic change from $(\pi,\pi,\pi)$ and $(0,0,\pi)$ to the PM phase, the change for $(0,\pi,\pi)$ [see Fig.~\ref{fig:fig3}(c)] is less pronounced, but still accurately resolvable. $T_{\text{N\'eel}}$ along different cuts of fixed $J_3$ as a function of $J_2$ is shown in Fig.~\ref{fig:fig3}(e). For cuts where there is a phase transition between two different magnetic orders (i.e., $J_3/J_1=0$ and $0.45$), we observe a clean kink in $T_{\text{N\'eel}}$ at the transition point. While we find a monotonous decrease of $T_{\text{N\'eel}}$ from the $(\pi,\pi,\pi)$ phase into the PM phase ($J_3/J_1=0.15$ and $0.30$), we observe a cusp feature upon reentering magnetically ordered phases for larger $J_2$.

Due to the sole availability of two-spin correlators, any further investigation of the PM phase is limited within PFFRG. As already successful for 2D PFFRG~\cite{Suttner-2014}, however, we can analyze the dimer response function to learn about propensities for translation symmetry breaking in the PM phase. The most basic valence bond crystal (VBC) candidates for the simple cubic lattice are depicted in Figs.~\ref{fig:fig4}(a)-\ref{fig:fig4}(c). Biasing the RG flow by slightly strengthening ($J_1\to J_1+\delta$) or weakening ($J_1\to J_1-\delta$) the nearest-neighbor couplings according to those patterns, we keep track of the dimer susceptibility which we define by

\begin{equation}
\chi^\text{dimer}=\frac{J_1}{\delta}\frac{C_+^\Lambda - C_-^\Lambda}{C_+^\Lambda + C_-^\Lambda}\;.\label{dimer_susceptibility}
\end{equation}

Here, $C^\Lambda_+$ ($C^\Lambda_-$) is the static spin-spin correlator on strong (weak) bonds. Note that Eq.~(\ref{dimer_susceptibility}) is normalized such that $\chi^\text{dimer}>1$ ($\chi^\text{dimer}<1$) corresponds to an enhancement (rejection) of the perturbation during the RG flow. As shown in Fig.~\ref{fig:fig4}(d), the columnar and plaquette VBC pattern is hardly amplified through the RG flow, and almost does not change throughout the PM phase. A similar behavior is observed for the cubic VBC, which, however, shows a moderate enhancement for decreasing $J_3$ within the PM phase. While a rigorous conclusion cannot be drawn at this stage, the similar magnitude of all three dimer susceptibilities leads us to conclude that no such VBC order is to be expected for the PM phase. Furthermore, compared to the antiferromagnetic $J_1$-$J_2$ model on the 2D square lattice, where the nature of the paramagnetic phase is still debated, dimer responses generally turn out to be smaller for the 3D cubic lattice~\cite{Reuther-2010}.

%%%%%%%%%%%%%%%%%%%%%%%%%%%%%%%%%%%%%%%%%%%%
\floatsetup[table]{capposition=top}
\begin{table}
\centering
\begin{tabular}{llllll}
 \hline \hline
      \multicolumn{1}{l}{$J_{3}/J_{1}$}
    & \multicolumn{1}{c}{$0$}
    & \multicolumn{1}{c}{$0.20$}
    & \multicolumn{1}{c}{$0.40$} 
    & \multicolumn{1}{c}{$0.60$} 
    & \multicolumn{1}{c}{$0.80$}  \\ \hline
       
\multirow{1}{*}{PFFRG} & $1.05(5)$ & $1.43(7)$ & $1.67(8)$ & $1.94(9)$ & $2.36(10)$    \\

\multirow{1}{*}{QMC} & $0.946(1)^{*}$ & $1.371(1)$ & $1.7675(10)$ & $2.143(1)$ & $2.5039(5)$ \\ \hline \hline

\end{tabular}
\caption{For the $J_{1}$-$J_{3}$ model, the N\'eel temperature $T_{\text{N\'eel}}/J_{1}$ as obtained from PFFRG and QMC is given. The result marked by an asterisk is from Ref.~\cite{PhysRevLett.80.5196}.}
\label{tab:t-neel}
\end{table}
%%%%%%%%%%%%%%%%%%%%%%%%%%%%%%%%%%%%%%%%%%%%

{\it Benchmark against other methods.} Wherever applicable, the qualitative and quantitative features of the phase diagram we find by PFFRG [see Fig.~\ref{fig:fig1}(a)] tend to agree with previous works. For $J_3=0$, we find the transition between $(\pi,\pi,\pi)$ and $(0,\pi,\pi)$ to occur at $J_{2}^{\rm c}=0.30(1)$, which validates earlier spin wave analysis of the $J_1$-$J_2$ model~\cite{Irkhin-1992,nonlin,PhysRevB.93.041106}. A variational cluster approach (VCA) study on a Hubbard model whose strong coupling limit maps onto Eq.~\eqref{ham} has found similar features in the magnetic phase diagram~\cite{PhysRevB.93.041106}. Recent efforts to extend the coupled cluster method (CCM) to 3D have met with some success~\cite{Schmalfuss-2006}. However, its application to the spin-$\frac{1}{2}$ Heisenberg $J_{1}$-$J_{2}$ antiferromagnet on the simple-cubic lattice predicts the appearance of a tiny paramagnetic phase in the vicinity of $J_{2}/J_{1}\approx 0.275$~\cite{PhysRevB.93.235123} similar to the finding within linear spin-wave approximation, but in contradiction to ours.

Regarding the benchmarking of $T_{\text{N\'eel}}$, we can only resort to quantum Monte Carlo (QMC) calculations in limits where there is no sign problem, such as the $J_1-J_3$ SC-AFM. In particular, for the $J_{1}$-only model, inaccurate early results have been corrected due to improved stochastic series expansion employed in QMC to yield $T_{\text{N\'eel}}/J_1 = 0.946 \pm 0.001$~\cite{PhysRevB.43.5950,PhysRevLett.80.5196}. The data point is depicted as a black asterisk in Fig.~\ref{fig:fig3}(e), to be compared with the PFFRG result of $T_{\text{N\'eel}}/J_1 = 1.05 \pm 0.05$, which reveals a remarkable agreement between QMC and 3D PFFRG. It is worth noting that the mean-field N\'eel temperature $T_{\text{N\'eel}}/J_1=1.5$ is still significantly larger. This shows that despite the mean-field character of the PFFRG, quantum fluctuations reducing $T_{\text{N\'eel}}$ are correctly built in. To further appreciate the quantitative accuracy of our approach we have carried out QMC simulations employing the ALPS/LOOPER program~\cite{ALPS,*Looper,*PhysRevLett.87.047203,*Albuquerque20071187,*1742-5468-2011-05-P05001}. By means of finite-size-scaling analysis of the Binder ratio for lattice sizes $8\leqslant L \leqslant 16$ we have determined $T_{\text{N\'eel}}$ for $J_{3}/J_{1}\leqslant 0.8$. A comparison with the PFFRG estimates (see Table~\ref{tab:t-neel}) reveals very good agreement. Together with various ways to obtain the Curie-Weiss temperature $T_\text{CW}$ as a function of Hamiltonian parameters, the 3D PFFRG thus provides a suitable way to compute the frustration parameter $f=T_\text{CW}/T_{\text{N\'eel}}$ in theoretical model calculations.

Aside from quantitatively reliable estimates of $T_{\text{N\'eel}}$ in frustrated magnets, the most applicable asset of 3D PFFRG appears to be the spin-spin correlation profile for large system sizes of several thousand sites. There is, however, no other quantum method in this range which one could compare with. Exact diagonalization methods and DMRG, for which quantitative comparisons with good agreement have at
least been possible for 2D PFFRG~\cite{Reuther-2011b,Suttner-2014}, cannot be fruitfully applied in 3D. Ultimately, the quantitative analysis of neutron scattering data with the help of 3D PFFRG susceptibility profiles will determine the degree of its utility. 

{\it Conclusion and outlook.} A methodological development in the field of frustrated quantum magnetism such as 3D PFFRG lends itself to various applications and investigations. Of particular interest are lattices where quantum fluctuations are expected to play a pivotal role such as the pyrochlore lattice for which quantum corrections might drive the system into new quantum states of matter~\cite{PhysRevLett.80.2933,PhysRevLett.112.207202}. Furthermore, other systems might prove promising where there are limits in which comparisons can be made against exact paramagnetic solutions, such as for the Kitaev model on three-coordinated lattices~\cite{PhysRevB.79.024426,PhysRevB.93.085101}. Finally, explicit material candidates on the hyperhoneycomb~\cite{PhysRevLett.114.077202} and hyperkagome~\cite{PhysRevLett.99.137207,PhysRevB.94.064416} lattice are ideal test beds to establish 3D PFFRG as a useful novel tool in frustrated magnetism.

%%%%%%%%%%%%%%%%%%%%%%%%%%%%%%%%%%%%%%%%%%%%%

{\it Acknowledgments} We thank F. F. Assaad for discussions on QMC. The work was supported by the European Research Council through ERC-StG-TOPOLECTRICS-Thomale-336012. Y.I. and R.T. thank the DFG (Deutsche Forschungsgemeinschaft) for financial support through SFB 1170. F.P.T. is supported by the DFG through DFG-FOR 1162  (AS 120/6-2). S.R. is supported by the DFG through SFB 1143 and by the Helmholtz association through VI-521. J.R. was supported by the Freie Universit\"at Berlin within the Excellence Initiative of the German Research Foundation. We gratefully acknowledge the Gauss Centre for Supercomputing e.V. for funding this project by providing computing time on the GCS Supercomputer SuperMUC at Leibniz Supercomputing Centre (LRZ). We thank the Center for Information Services and High Performance Computing (ZIH) at TU Dresden for generous allocations of computer time.

\end{document}